\documentclass[12pt]{iopart}
\usepackage{graphicx,graphics}

\expandafter\let\csname equation*\endcsname\relax
\expandafter\let\csname endequation*\endcsname\relax

\usepackage{subcaption, amsmath}

\usepackage{gensymb}
\usepackage{booktabs, comment}
\usepackage{multirow}
\usepackage[utf8]{inputenc}
\usepackage{lscape}
\usepackage{float}
\usepackage[dvipsnames]{xcolor}
\usepackage{comment}
\usepackage{tikz}
\usepackage{physics}
\usepackage{hyperref}
\usepackage{cite}
\newcommand*\circled[1]{\tikz[baseline=(char.base)]{
            \node[shape=circle,draw,inner sep=2pt] (char) {#1};}}
\newcommand{\SN}[1]
{{\color{black}#1}} 
\newcommand{\JRS}[1]{{\color{black}#1}} 
\newcommand{\NI}[1]{{\color{black}#1}} 
\usepackage{xcolor} 

\definecolor{darkgreen}{rgb}{0,0.5,0} 
\newcommand{\trlee}[1]{{\color{black}#1}} 

\definecolor{lime}{HTML}{A6CE39}
\DeclareRobustCommand{\orcidicon}{
	\begin{tikzpicture}
	\draw[lime, fill=lime] (0,0) 
	circle [radius=0.16] 
	node[white] {{\fontfamily{qag}\selectfont \tiny ID}};
	\draw[white, fill=white] (-0.0625,0.095) 
	circle [radius=0.007];
	\end{tikzpicture}
	\hspace{-2mm}
}

\foreach \x in {A, ..., Z}{\expandafter\xdef\csname orcid\x\endcsname{\noexpand\href{https://orcid.org/\csname orcidauthor\x\endcsname}
			{\noexpand\orcidicon}}
}

\begin{document}

\title[]{Polarisation and Temperature Dependence of Er$^{3+}$:CaWO$_4$ -- Towards a Solid-State Rare-Earth Ion-Doped Quantum Memory} 

\newcommand{\orcidauthorA}{0000-0002-3437-1200} 
\newcommand{\orcidauthorB}{0009-0000-3970-0803} 
\newcommand{\orcidauthorC}{0000-0002-9174-8529} 
\newcommand{\orcidauthorD}{0000-0001-6601-1907} 
\newcommand{\orcidauthorE}{0000-0001-8732-3352} 
\newcommand{\orcidauthorF}{0000-0003-0654-0393} 
\newcommand{\orcidauthorG}{0000-0002-4421-601X} 
\newcommand{\orcidauthorH}{0000-0002-9673-3916} 
\newcommand{\orcidauthorI}{0000-0002-0842-6984} 

\author{Mikhael T. Sayat$^{1,2,3*\orcidA} $, Trevor R. Lee$^{1,2\orcidB}$, Suchit Negi$^{1,2,4\orcidC}$, Naoya Iwahara$^5\orcidD$, In Cheol Seo$^{1,6\orcidE}$, Yung Chuen Tan$^{1,6\orcidF}$, Ping Koy Lam$^{1,2,3\orcidG}$, Young-Wook Cho$^{1,2,3\#\orcidH}$, Jian-Rui Soh$^{1,2,3\dagger\orcidI}$}

\address{1. Quantum Innovation Centre (Q.InC), Agency for Science, Technology and Research (A*STAR), 2 Fusionopolis Way, Innovis \#08-03, Singapore 138634, Singapore}
\address{2. Institute of Materials Research and Engineering (IMRE), Agency for Science Technology and Research (A*STAR), 2 Fusionopolis Way, Innovis \#08-03, Singapore 138634, Republic of Singapore}
\address{3. Centre for Quantum Technologies, National University of Singapore, 3 Science Drive 2, Singapore 117543, Singapore}
\address{4. Department of Physics, Faculty of Science, National University of Singapore, Science Drive 3, Singapore 117551, Singapore}
\address{5. Graduate School of Engineering, Chiba University, 1-33 Yayoi-cho, Inage-ku, Chiba-shi, Chiba 263-8522, Japan}
\address{6. National Metrology Centre (NMC), Agency for Science, Technology and Research (A*STAR), Singapore 637145}

\ead{*mikhael\_sayat@imre.a-star.edu.sg, \#cho\_youngwook@imre.a-star.edu.sg, $\dagger$soh\_jian\_rui@imre.a-star.edu.sg}
\vspace{10pt}
\begin{indented}
\item[] July 2025
\end{indented}

\begin{abstract}
In the endeavour of developing quantum memories, Er$^{3+}$:CaWO$_4$ has emerged as a promising rare-earth ion-doped (REID) crystal platform due to its long optical coherence times and compatibility with the 1550\,nm telecommunications band. This work investigates the effects of polarisation and temperature on the absorption strength, central wavelength, and linewidth of the $Z_1\to Y_1$ and $Z_1\to Y_2$ optical transitions, with light incident along the crystal $a$ and $c$ axes. It is found that the $Z_1\to Y_1$ transition at 1532.6\,nm with the incident laser along the $c$-axis at cryogenic temperatures ($\sim$\,3\,K) is particularly favourable. The transition exhibits a stable central wavelength, narrower linewidth, polarisation independence, larger absorption cross-section, and lies within the C-band --attributes that make it highly suitable for quantum memory applications.
\end{abstract}

%
\vspace{2pc}
\noindent{\it Keywords}: quantum memory, rare-earth ion-doped crystal, absorption spectra, optical transition, telecommunication
%
%
%
%

\section{Introduction} 
The recent advancement of quantum technologies has made the grand goal of a global quantum network more feasible, featuring advances in computation, communication, sensing, and fundamental physics \cite{simon2017towards,mol2023quantum}. This would usher in new quantum-enhanced technologies such as repeaters and relays \cite{briegel1998quantum,munro2015inside,liorni2021quantum,azuma2023quantum}, computers \cite{ladd2010quantum,preskill2018quantum,rietsche2022quantum}, sensors \cite{zaiser2016enhancing,yang2019memory}, and cryptography \cite{biham1996quantum,gundougan2021proposal,bhaskar2020experimental,wang2022field}.

An overarching technology in this advancement is the quantum memory, which can store and retrieve information encoded in quantum states \cite{lvovsky2009optical}. The performance of quantum memories can be quantified with three figures of merit:  bandwidth -- allowable \trlee{repetition rate of the information storage}, efficiency -- probability of successfully retrieving the stored quantum state, and storage time -- period of time the quantum state is retained within the quantum memory \cite{mol2023quantum, dajczgewand2015optical}. A variety of platforms, with different strengths and weaknesses in these metrics, have been developed. Examples include warm vapour cells \cite{pinel2013gradient,hosseini2011unconditional}, cold atoms \cite{cho2016highly}, and single atoms/ions defects \cite{hedges2010efficient,heshami2016quantum,jing2024approaching}. 

A promising platform is rare-earth-ion-doped crystals, which have long storage and coherence times \cite{nilsson2005coherent,liu2025millisecond}, narrow optical transitions at cryogenic temperatures \cite{GOLDNER20151} for controlling the quantum states, and wide bandwidths \cite{liu2025millisecond} to accommodate for a variety of wavelengths, and thus diverse applications. Significant advancements in quantum memories using rare-earth-ion-doped crystals include the demonstration of the coherent storage of light for 1 hour \cite{ma2021one}, and observed coherence times of 6 hours \cite{zhong2015optically} and 18 hours \cite{wang2025nuclear}, in Eu$^{3+}$:Y$_2$SiO$_5$. In addition, the storage of qubits has been demonstrated for 1.021\,ms \cite{liu2025millisecond} and 20\,ms \cite{ortu2022storage}. Entangled photons have also been heralded in quantum memories with a bandwidth of 1~GHz using  Nd$^{3+}$:YVO$_4$ \cite{liu2021heralded}, and have also been interfaced with Ti:Tm:LiNbO$_3$ waveguides with a bandwidth of 5~GHz \cite{saglamyurek2011broadband}. Proposals have also been made with host crystals doped with Kramer's ions ($\mathrm{Er^{3+}}$, $\mathrm{Nd^{3+}}$, $\mathrm{Yb^{3+}}$) with bandwidths greater than 10\,GHz \cite{vivoli2013high}. 

A propitious rare-earth-ion-doped (REID) crystal is erbium-doped calcium tungstate Er$^{3+}$:CaWO$_4$, where coherence times of 1.3\,ms \cite{ranvcic2022electron} and 23\,ms \cite{le2021twenty} have been achieved using microwave resonators at cryogenic temperatures. The host crystal, calcium tungstate (CaWO$_4$), has a low spin bath since most nuclei in the crystal have no nuclear spin, which could lead to long coherence times \cite{le2021twenty}. This is in contrast with the Y$_2$SiO$_5$ host crystal, which has been the conventional host for rare-earth ions, where the only stable isotope of yttrium (Y) is nuclear active. In addition, the rare earth ion dopant, erbium (Er$^{3+}$), has optical transitions near 1550\,nm (C-band) suitable for communication purposes and easier integration into existing telecommunication networks \cite{ourari2023indistinguishable,uysal2025spin}. This makes Er$^{3+}$:CaWO$_4$ an attractive and promising REID crystal for developing quantum memories. However, studies have so far have predominantly been focused on the microwave regime leaving a knowledge gap in the optical regime; in particular, the operating regime of Er$^{3+}$:CaWO$_4$ for quantum memory applications. In this paper, the effects of linear polarisation and temperature on the optical transition strengths of Er$^{3+}$:CaWO$_4$ are investigated to address the suitability of storing quantum information. The paper is structured as follows: Section 2 presents the electronic and crystal structure of Er$^{3+}$:CaWO$_4$, Section 3 presents the experiment and method for investigating the polarisation and temperature dependence, Section 4 presents the results, Section 5 provides a discussion and future work, and Section 6 concludes the study.




\section{Er$^{3+}$:CaWO$_4$} 

\subsection{\JRS{Crystal Structure}}
\begin{figure}[ht!]
\begin{subfigure}[t]{0.24\textwidth}
\centering
\includegraphics[width=\linewidth]{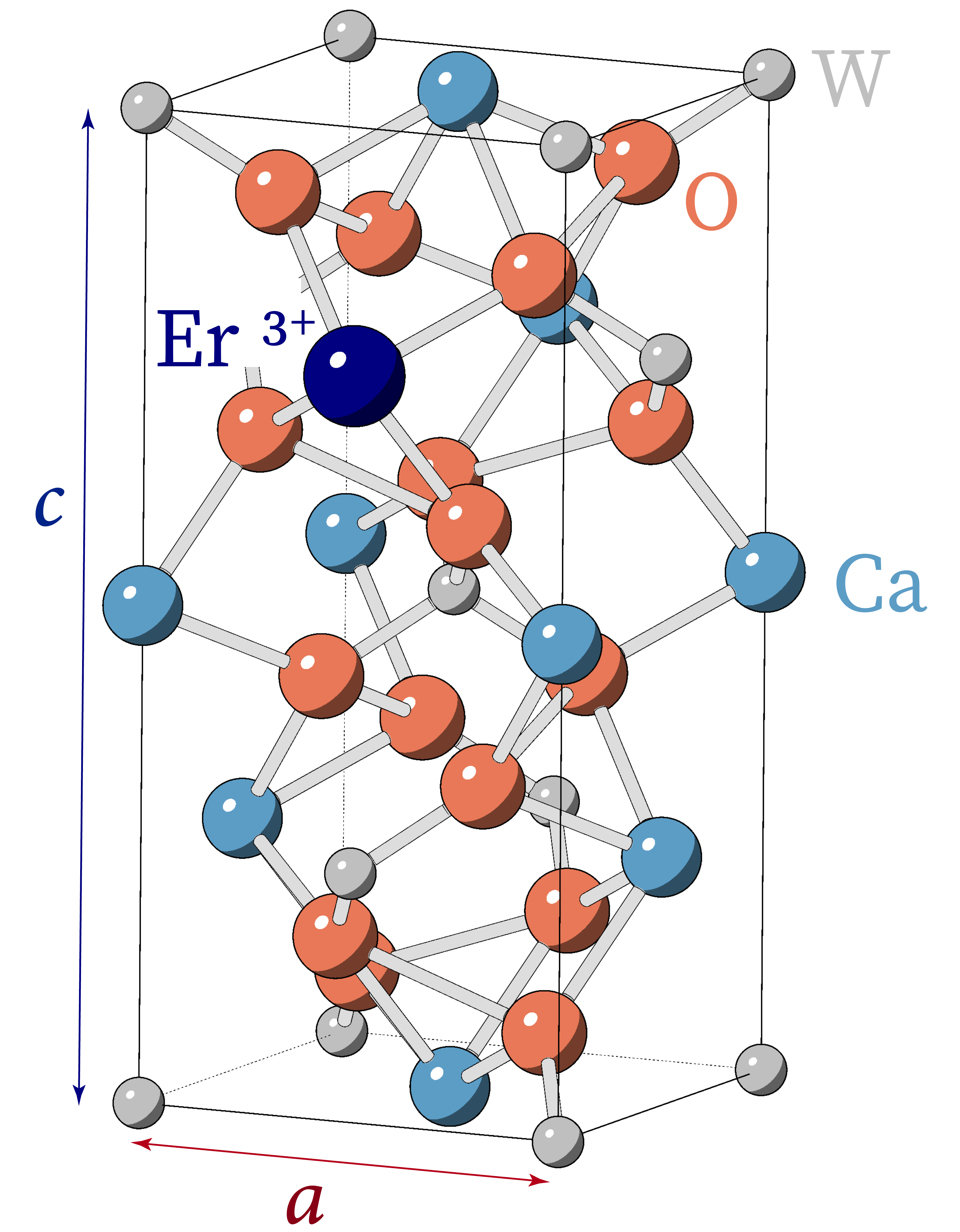}
       \caption{}
    \end{subfigure}
    ~ 
\begin{subfigure}[t]{0.38\textwidth}
        \centering
        \includegraphics[width=\linewidth]{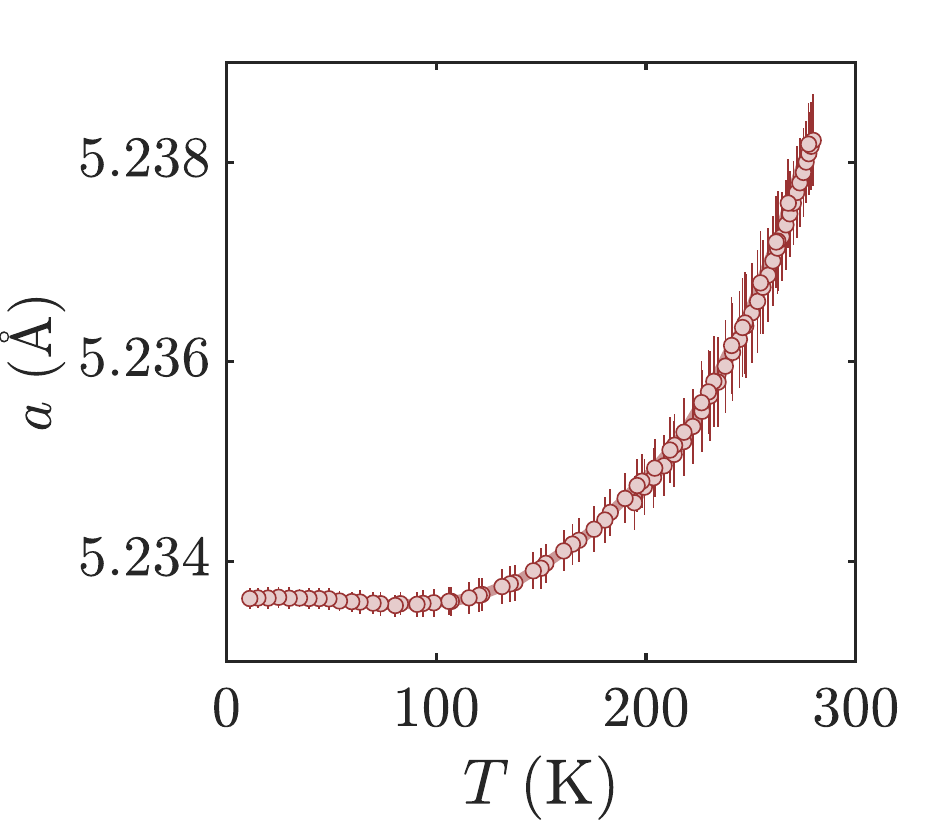}
       \caption{}
    \end{subfigure}
    ~ 
    \begin{subfigure}[t]{0.38\textwidth}
        \centering
        \includegraphics[width=\linewidth]{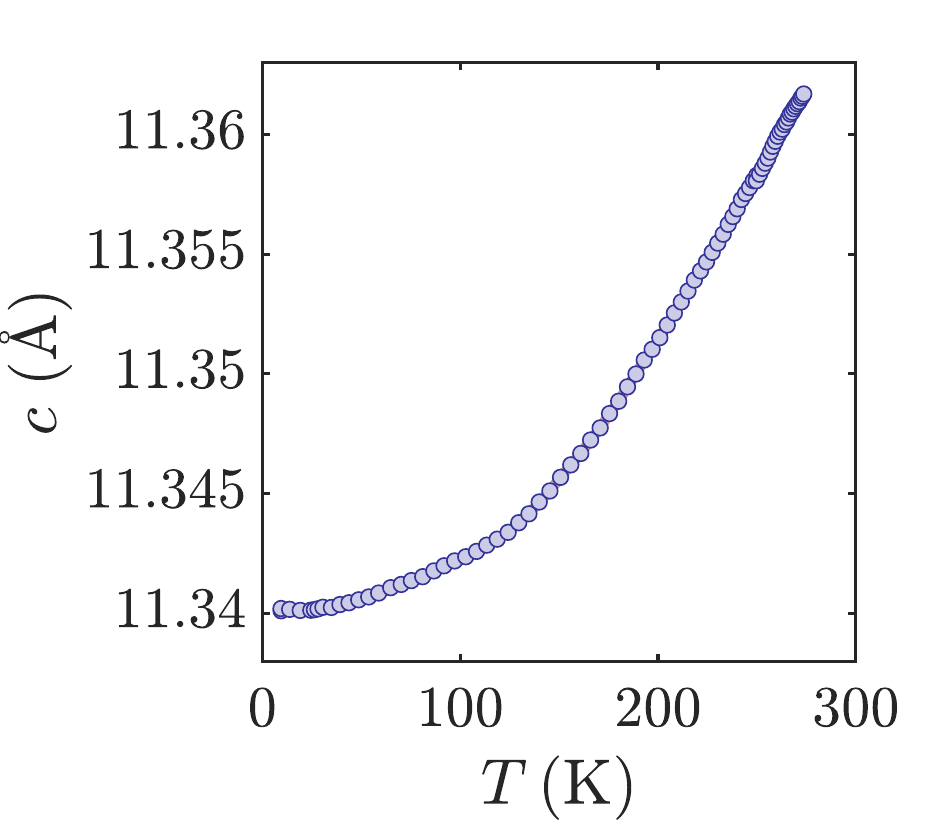}
        \caption{}
    \end{subfigure}
    \caption{\label{fig:Crystal_structure_temp} \JRS{(a) The tetragonal unit cell CaWO$_4$ can be described by the $I4_1/a$ space group. The Er$^{3+}$ (blue) dopants replace the Ca ions, which reside on the $4b$ Wyckoff position [$I4_1/a$; origin choice \#1]. (b), (c) Temperature dependence of the (4,0,0) and (0,0,8) reflection of single crystalline CaWO$_4$, measured on the BL-3A beamline.}}
\end{figure}

\SN{The host material calcium tungstate ($\mathrm{CaWO_4}$) belongs to the scheelite family with a tetragonal \JRS{crystal} structure (space group I4$_{1}$/a, No. 88) \cite{becker2024spectroscopic}. In this work, the setting \#1 of the $I4_1/a$ space group, in which the calcium (Ca), tungsten (W), and oxygen (O) atoms reside on the $4b$, $4a$ and $16f$ Wyckoff sites respectively [Fig.~\ref{fig:Crystal_structure_temp}(a)]}, was adopted. \JRS{Given that the unit cell of CaWO$_4$ is tetragonal, the crystal $a$ and $b$ axes are symmetry equivalent, related by a 4-fold ($4_1$) screw axes along the crystal $c$ axes. As such, in the subsequent discussion, it is sufficient to consider the system only along the $a$ and $c$ axes, without loss of generality.} 

\JRS{The CaWO$_4$ single crystals used were grown via the hybrid flow-zone-Czochralski method (SurfaceNet GmbH) with an Er$^{3+}$ dopant concentration of 50\,ppm. These erbium dopants occupy the Ca sites, as depicted in Fig.~\ref{fig:Crystal_structure_temp}(a). Single-crystal x-ray diffraction with an incident x-ray wavelength of $\lambda$=1.5406\,\AA{} (Cu $K_\alpha$) using the 6-circle diffractometer (D8 VENTURE, Bruker) was performed to determine the cell parameters of the host crystal, CaWO$_4$, at room temperature. It was found that the crystal $a$ and $c$ cell parameters were 5.2398(4)\,\AA{} and 11.3622(7)\,\AA, respectively, which is in good agreement with those obtained in Ref.~\cite{cif2023}. 

To understand how the crystal cell parameters change as a function of temperature, high-resolution single-crystal x-ray diffraction was performed on the BL-3A beamline at the KEK photon factory. The measurements were performed in the vertical scattering geometry, with the incident photon energy fixed at 12\,keV. A silicon drift detector was used to detect the scattered x-rays, \trlee{while a closed-cycle He-4 cryostat provided temperature regulation.} Figure~\ref{fig:Crystal_structure_temp}(b), (c) plots the temperature dependence of the $a$ and $c$ cell parameters, based on the reciprocal space location of the structurally-allowed $\textbf{Q}$=(4,0,0) and $\textbf{Q}$=(0,0,8) reflections of CaWO$_4$, respectively. It was found that the $a$ and $c$ cell parameters do not change appreciably, below $\sim$10\,K. This means that the operating temperature of the CaWO$_4$ host crystal should ideally be below 10\,K to avoid thermal-expansion-induced misalignment of the laser with respect to the host crystal.}

\subsection{\SN{Electronic Structure of CaWO$_4$}}
\JRS{To understand the electronic structure of the host crystal at low temperatures, \textit{ab-initio} density functional theory (DFT) calculations of CaWO$_4$ were performed using the Quantum Espresso package \cite{giannozzi2017}. A plane wave basis set together with optimised norm-conserving Vanderbilt (ONCV) pseudopotentials~\cite{oncv2015} was adopted to account for the core electrons. The kinetic energy cutoff of 50\,Ry was used together with the exchange-correlation functional of Perdew, Burke and Ernzerhof (PBE) \cite{pbe1996}. The Brillouin zone was sampled using a $\Gamma$-centred 6$\times$6$\times$6 Monkhorst-Pack (MP) grid \cite{mpgrid} for all calculations, together with the Broyden-Fletcher-Goldfarb-Shanno algorithm \cite{BFGS} for geometry optimisation. 
}

\begin{figure}[hb!]
    \centering
    \includegraphics[width=\linewidth]{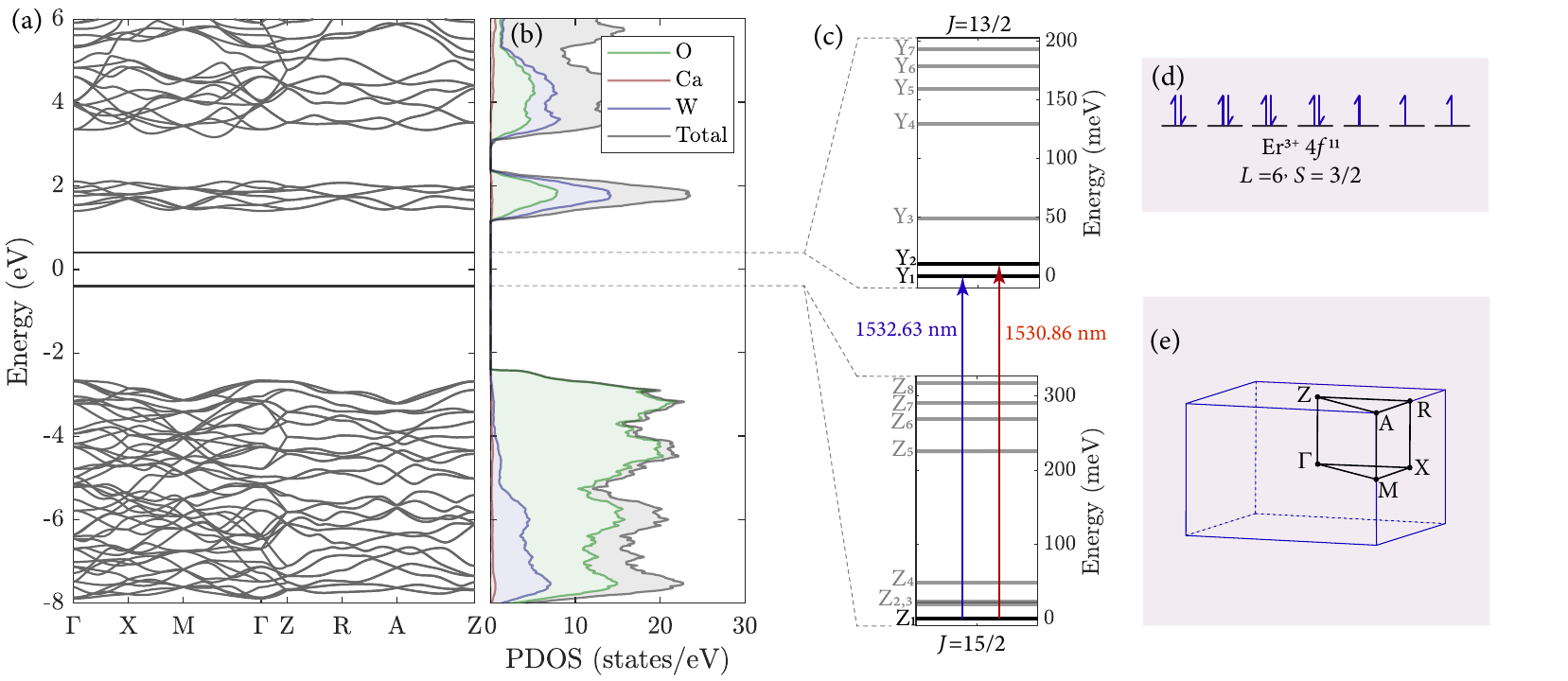}
    \caption{CaWO$_4$ electronic structure with Fermi level set to 0\,eV. (a) Band-gap and (b) projected density of states with contributions from Ca, O (\textit{s,p}) and W (\textit{s,p,d}) orbitals. Er dopant state appears as a dotted line within the gap. (c) Calculated crystal electric field splitting of the $J$=$\frac{15}{2}$ and $\frac{13}{2}$ manifold of Er$^{3+}$. (d) The $4f^{11}$ configuration of Er$^{3+}$, with orbital and spin angular momentum of $L$=6 and $S$=$\frac{3}{2}$. (e) Tetragonal Brillouin zone of CaWO$_4$.}
    \label{fig:CaWO4_electroni}
\end{figure}

\SN{From the structural relaxation of CaWO$_4$, the relaxed cell lattice parameters obtained were $a$ = $b$ = 5.039\,\AA, $c$ = 10.768\,\AA{} and $\alpha$ = $\beta$ = $\gamma$ = 90$^\circ$. This is within 4\% of the measured cell parameters, which is the common acceptable discrepancy between experiment and DFT calculations.}
\SN{Figure~\ref{fig:CaWO4_electroni}(a) plots the electronic band structure of CaWO$_{4}$ along high-symmetry paths~\cite{setya2010} in the tetragonal Brillouin zone [See Fig.~\ref{fig:CaWO4_electroni}(e)]. Furthermore, the atomic orbitals were projected onto the electronic density of states (DOS), as shown in Fig.~\ref{fig:CaWO4_electroni}(b) with contributions from Ca, O (\textit{s,p}) and W (\textit{s,p,d}) orbitals. 
The DFT calculations predict a absence of states in the vicinity of the Fermi energy with a large bandgap of 4.06\,eV, confirming the insulating nature and near-infrared \JRS{transparency} of CaWO$_4$, making it a suitable host material for Er dopants \JRS{with transitions in the telecommunications C-band}.}

\newpage
\subsection{\JRS{Energy levels of Er$^{3+}$ dopants}}
\JRS{Having discussed the electronic structure of the CaWO$_4$ host crystal, it is now important to consider the electronic transitions of the dopant, Er$^{3+}$. The three unpaired electrons in the $4f^{11}$ configuration of the Er$^{3+}$ ion carry a total orbital angular momentum $L=6$ and spin angular momentum $S=\frac{3}{2}$. Due to spin-orbit coupling, these combine to give an effective spin-orbit coupled states $J=\frac{15}{2}$, $\frac{13}{2}$, $\frac{11}{2}$ and $\frac{9}{2}$, following Hund's rules. In particular, the transitions of focus are between the spin-orbit coupled $J=\frac{15}{2}$ and $J=\frac{13}{2}$ manifolds of Er$^{3+}$, which possess optical transitions near the telecommunications wavelength of 1550\,nm. As described earlier, the Er$^{3+}$ dopants occupy interstitial sites within the host crystal lattice, substituting for Ca$^{2+}$ ions, as illustrated in Fig.~\ref{fig:Crystal_structure_temp}(a). This substitution breaks the spherical symmetry of the free ion and lifts the 
degeneracy of the spin–orbit manifolds due to the crystal electric field generated by the surrounding ligands. As a result, the $J
=\frac{15}{2}$ ground state splits into eight crystal-field levels, denoted $Z_i$ (where
$i=$ 1 to 8), and the $J=\frac{13}{2}$ excited state splits into seven levels, denoted $Y_j$ (where $j=$ 1 to 7). 

\NI{The crystal-field-induced splittings within the $J=\frac{15}{2}$ and $J=\frac{13}{2}$ manifolds were calculated using a crystal field Hamiltonian $\hat{H}^J_\text{CF}$ expressed in terms of irreducible tensor operators (ITOs) acting on the $4f^{11}$ configuration of Er$^{3+}$ 
(See Supplemental Materials for details). The assumptions made were that the Er$^{3+}$ site has $S_4$ point group symmetry, the crystal field originates from a single electron interaction, and neglecting any mixing between different $J$ multiplet states. Under these assumptions, the crystal field Hamiltonian within a given ${}^4I_{J}$ multiplet takes the form 
\begin{align}
 \hat{H}^J_\text{CF} &= 
 B_{20}^J \hat{\mathcal{T}}_{20}^J + 
 B_{40}^J \hat{\mathcal{T}}_{40}^J + 
 B_{44}^J \left( \hat{\mathcal{T}}_{44}^J + \hat{\mathcal{T}}_{4,-4}^J \right)
 \nonumber\\
 &+
 B_{60}^J \hat{\mathcal{T}}_{60}^J + 
 B_{64}^J \left( \hat{\mathcal{T}}_{64}^J + \hat{\mathcal{T}}_{6,-4}^J \right),
 \label{Eq:HCF_152}
\end{align}
with ITOs $\hat{\mathcal{T}}_{kq}$ (see Ref. \cite{Santini2009} and Supplemental materials).
Only the terms consistent with $S_4$ symmetry are retained, namely, $B_{20}$, $B_{40}$, $B_{44}$, $B_{60}$, and $B_{64}$. 
Higher-order terms due to the inter-manifold couplings are neglected. 
}

Figure~\ref{fig:CaWO4_electroni}(c) plots the energy levels obtained from the calculations, which show the splitting in the $J=\frac{15}{2}$ and $J=\frac{13}{2}$ manifold. In the following section, the transitions considered were those between the ground state $Z_1$ into the $Y_1$ to $Y_7$ manifold, paying special attention to the $Z_1\to Y_1$ and $Z_1\to Y_2$ transitions as they are closest to 1550\,nm in wavelength.}

\section{Experimental Method} 

\begin{figure}[htp!]
    \centering
    \includegraphics[width=\linewidth]
    {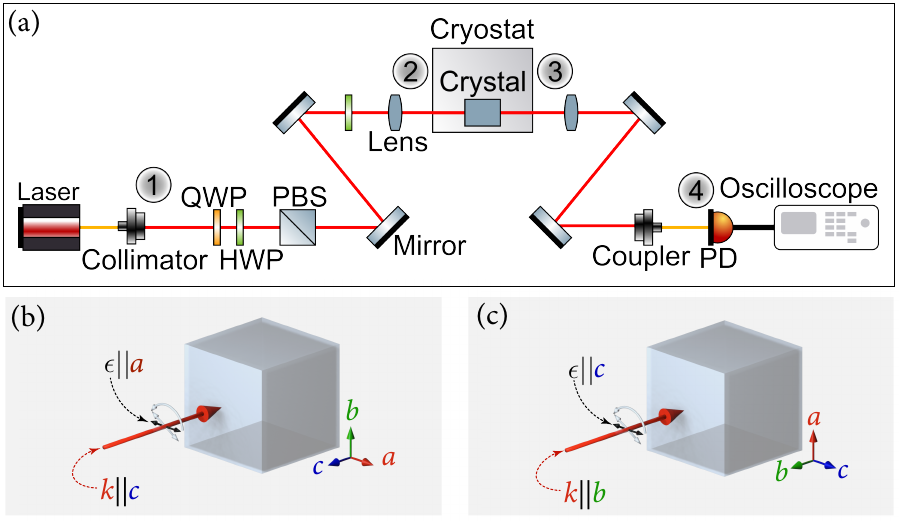}
    \caption{(a) Optical setup, QWP: quarter-wave plate. HWP: half-wave plate. PBS: polarising beam splitter. PD: photodetector. Positions \protect\circled{1}-\protect\circled{4} were used for power characterisation through the beam path.\JRS{The measurements were performed in two configurations with (b) $\textbf{k}||c$ and (c) $\textbf{k}||a$ (Note. $a$ and $b$ are equivalent for CaWO$_4$.). (b) For the $\textbf{k}||c$ configuration, the incident laser linear polarisation ($\epsilon$) was rotated about the $a$--$b$ plane, and the horizontal linear polarisation is defined at 0$^\circ$, with $\epsilon||a$. (c) For the $\textbf{k}||b$ configuration, the incident laser linear polarisation was rotated about the $a$--$c$ plane, and the horizontal linear polarisation is defined at 0$^\circ$, with $\epsilon||c$.}}
    \label{fig:FS_Setup}
\end{figure}

The experimental setup is shown in Fig.~\ref{fig:FS_Setup}(a). A tunable C-band narrowband laser (Santec TSL-570) is coupled to free-space using a fibre-collimator for subsequent optical manipulation. 
The laser passes through a linear polariser consisting of a quarter-wave plate (QWP), a half-wave plate (HWP), and a polarising beam splitter (PBS). The QWP and HWP were optimised to minimise the power of the reflected beam from the PBS, ensuring the laser is horizontally polarised. The second HWP was used for \trlee{linear} polarisation control. A lens was used to focus the light on a 5$\times$5$\times$5mm CaWO$_4$ single crystal with a Er$^{3+}$ dopant concentration of 50 ppm inside the \trlee{cryostat (Montana Instruments, CryoAdvance-50)}. The laser is then coupled to a multimode fibre, \trlee{measured by an InGaAs photodetector (PD) (Thorlabs Inc., PDA05CF2) connected to an oscilloscope (See Supplemental Materials for details on how the data was collected).}


The absorption, central wavelength, and linewidth of the $Z_1\to Y_1$ and $Z_1\to Y_2$ transitions were studied as a function of polarisation and temperature. As described earlier, these transitions were chosen as they are closest to the 1550\,nm telecommunication wavelength. \trlee{For power variation, the input power of the laser was varied while maintaining} horizontally polarised light and a temperature of 3.2\,K. \trlee{For polarisation variation,} the polarisation was varied using the second HWP [Fig.~\ref{fig:FS_Setup}] \trlee{while maintaining} a laser input power of 100~$\mu$W and a temperature of 3.2\,K. For temperature variation, the temperature within the cryostat was varied while maintaining horizontally polarised light and a laser input power of 100~$\mu$W. A laser input power of 100~$\mu$W was chosen for the polarisation and temperature dependence measurements because this order of magnitude for the laser input power produced a reasonable absorption with low noise (large SNR: signal-to-noise-ratio). 

\JRS{These polarisation- and temperature-dependent measurements were performed in two configurations with respect to the host crystal axes [Fig.~\ref{fig:FS_Setup}(b), (c)]. In the first configuration, the incident beam is oriented along the crystal $c$ axis, namely $\textbf{k}||c$, where $\textbf{k}$ is the incident laser wave vector. In this configuration, the linear polarization of the incident laser could be rotated within the $a$--$b$ plane. The horizontal linear polarisation was set along the crystal $a$ axis, (i.e. $\epsilon||a$\trlee{, where $\epsilon$ is the polarisation vector of the incident light}) [Fig.~\ref{fig:FS_Setup}(b)]. In the second configuration, where $\textbf{k}||a$, the incident linear polarisation of the laser could be rotated within the $a$--$c$ plane. In this configuration, the horizontal linear polarization is defined along the crystal $c$-axis (i.e. $\epsilon||c$) [Fig.~\ref{fig:FS_Setup}(c)].}

\section{Results} 
\begin{figure}[htp!]
    \centering
    \begin{subfigure}[t]{0.5\textwidth}
        \centering
        \includegraphics[width=\linewidth]{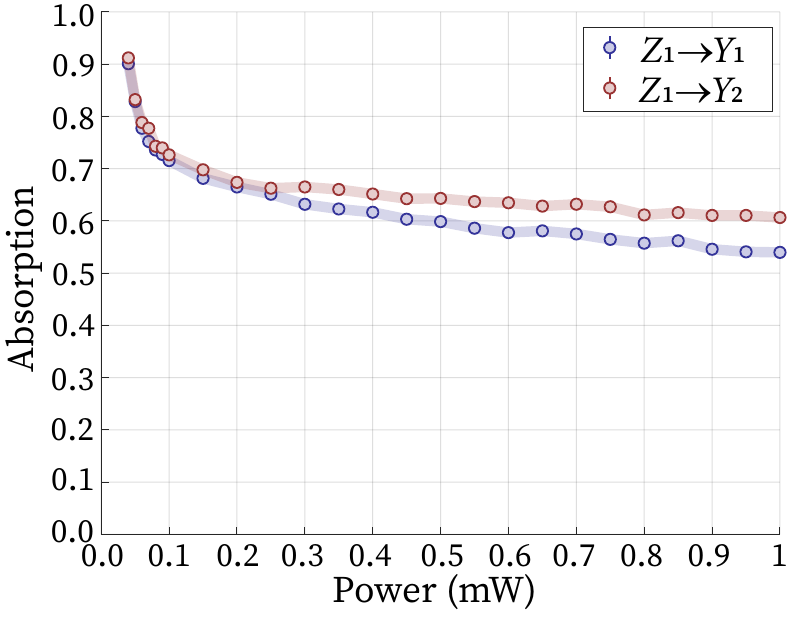}
        \caption{$\textbf{k}||c$ configuration with $\epsilon||a$.}
    \end{subfigure}%
    ~ 
    \begin{subfigure}[t]{0.5\textwidth}
        \centering
        \includegraphics[width=\linewidth]{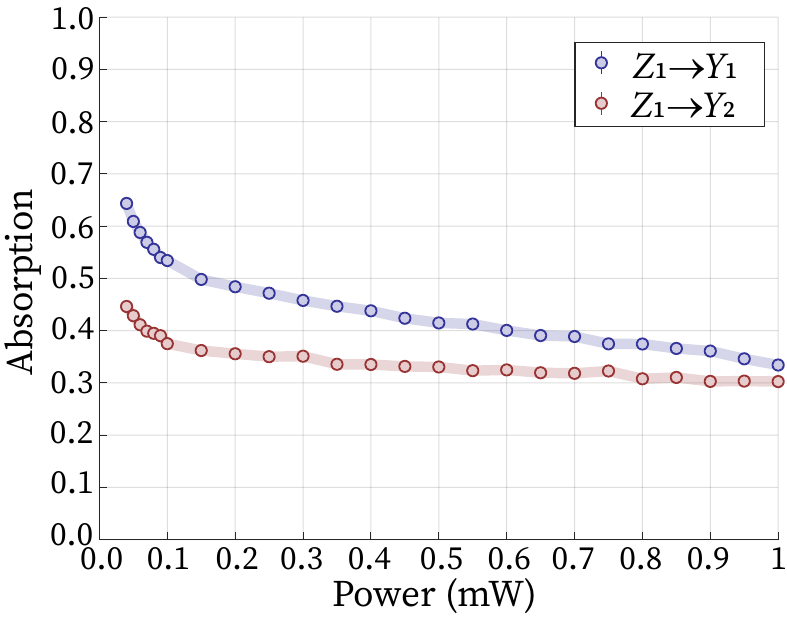}
        \caption{$\textbf{k}||a$ configuration with $\epsilon||c$.}
    \end{subfigure}%
    \caption{\label{fig:power}Absorption in the (a) $\textbf{k}||c$ configuration with $\epsilon||a$ and (b) $\textbf{k}||a$ configuration with $\epsilon||c$ as a function of input laser power.
    }
\end{figure}

Prior to the investigation on the effects of linear polarisation and temperature on the absorption, central wavelength, and linewidth for Er$^{3+}$:CaWO$_4$, the optical power was characterised. The effects of varying the input laser power are shown in Fig.~\ref{fig:power} \JRS{for the $\textbf{k}||c$ configuration with $\epsilon||a$ and for the $\textbf{k}||a$ configuration with $\epsilon||c$}.





The absorptions as a function of power for the $Z_1\to Y_1$ and $Z_1\to Y_2$ transitions show an exponential relationship. Figure~\ref{fig:power}(a) shows similar absorptions between the $Z_1\to Y_1$ and $Z_1\to Y_2$ transitions below a power of approximately 0.1\,mW, above which the $Z_1\to Y_2$ transition has slightly greater absorption than the $Z_1\to Y_1$ transition. This shows that the $Z_1\to Y_2$ transition saturates faster (curve flattens faster) in the $\textbf{k}||c$ configuration, meaning that there is a stronger interaction between the excitation laser and the Er$^{3+}$ ions in this transition. This faster saturation also means that the decay rate is large at this transition and the lifetime of the excited state is short -- translating to short storage times. This makes the $Z_1\to Y_1$ transition more favourable as it saturates more gradually and therefore has longer storage times.

Conversely, Figure~\ref{fig:power}(b) shows that the $Z_1\to Y_1$ transition overall has larger absorption than the $Z_1\to Y_2$ transition for powers below 1\,mW. It is shown once again, that the $Z_1\to Y_2$ transition saturates faster (curve flattens faster) indicating that in the $\textbf{k}||a$ configuration, this transition also has shorter storage times. In addition, it can be seen that the absorptions in the $\textbf{k}||a$ configuration are smaller than in the $\textbf{k}||c$ configuration. This indicates that the coupling in the $\textbf{k}||a$ configuration is weaker and could lead to smaller storage times.

Fig.~\ref{fig:power} shows that larger absorptions can be achieved for lower powers, with approximately 0.9 absorption for both transitions in the \JRS{$\textbf{k}||c$ configuration with $\epsilon||a$.} On the \JRS{other hand, in the $\textbf{k}||a$ configuration with $\epsilon||c$}, the $Z_1\to Y_2$ transition only reaches a measured absorption maximum of approximately 0.45 while the $Z_1\to Y_1$ transition reaches a maximum absorption of approximately 0.65. The behaviour of the two transitions in the  $\textbf{k}||c$ and $\textbf{k}||a$ configurations are different as shown by the the different levels and flattening of the curves in Fig. \ref{fig:power}. This implies that the behaviour of the crystal is dependent on the relative orientation of the laser beam with respect to the crystal axis. Moreover, this further implies that for a given configuration, the absorption can vary as a function of incident beam linear polarisation -- the subject of the consequent section.

\newpage
\subsection{Polarisation}
\begin{figure}[htp!]
    \centering
    \begin{subfigure}[t]{0.33\textwidth}
        \centering
        \includegraphics[width=\linewidth]{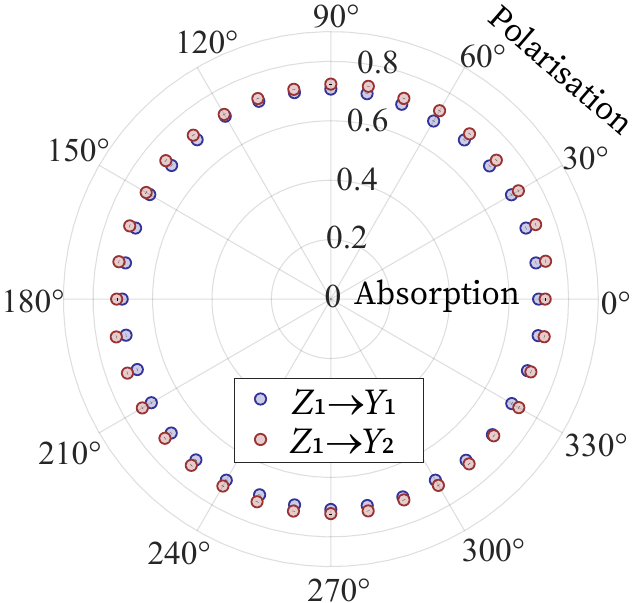}
        \caption{Absorption}
    \end{subfigure}%
    ~ 
    \begin{subfigure}[t]{0.33\textwidth}
        \centering
        \includegraphics[width=\linewidth]{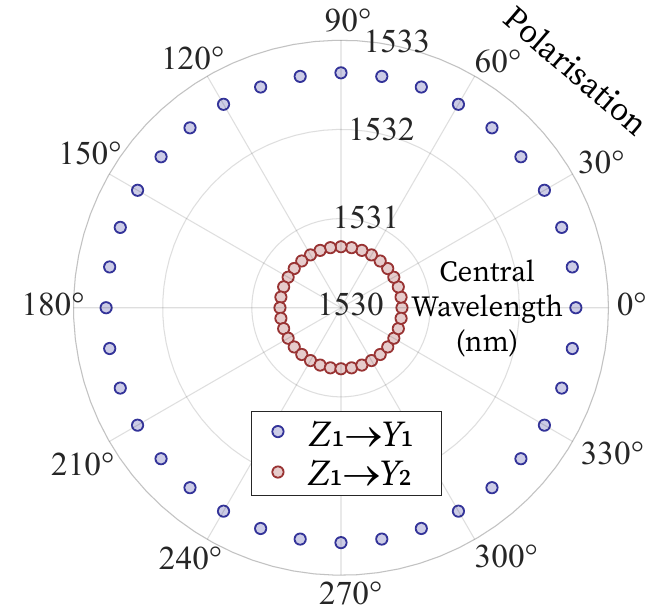}
        \caption{Central wavelength}
    \end{subfigure}%
    ~
    \begin{subfigure}[t]{0.33\textwidth}
        \centering
        \includegraphics[width=\linewidth]{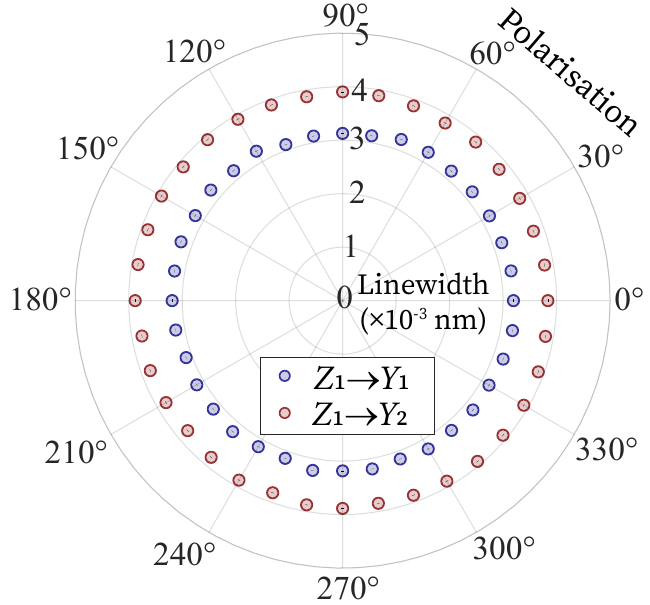}
        \caption{Linewidth}
    \end{subfigure}
    \caption{\label{fig:Polarisation_a}Absorption, central wavelength, and linewidth \JRS{in the $\textbf{k}||c$ configuration} as a function of polarisation angle. $0^\circ$ indicates horizontally polarised light with $\epsilon||a$.}
\end{figure}

\begin{figure}[htp!]
    \centering
    \begin{subfigure}[t]{0.33\textwidth}
        \centering
        \includegraphics[width=\linewidth]{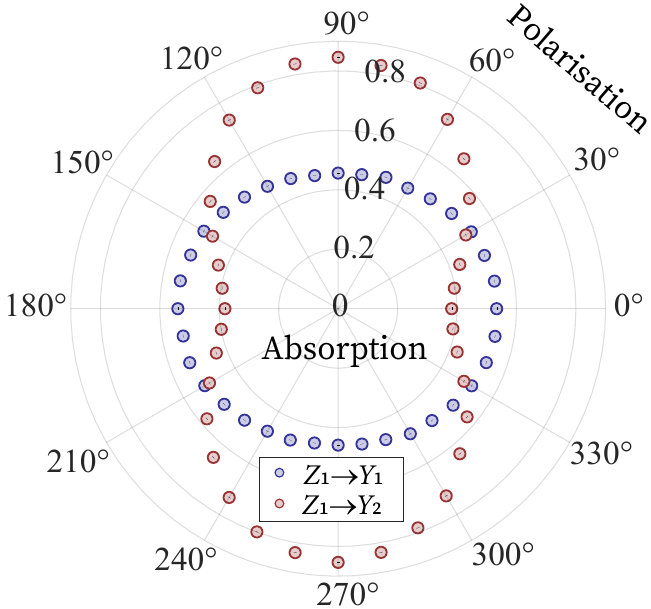}
        \caption{Absorption}
    \end{subfigure}%
    ~ 
    \begin{subfigure}[t]{0.33\textwidth}
        \centering
        \includegraphics[width=\linewidth]{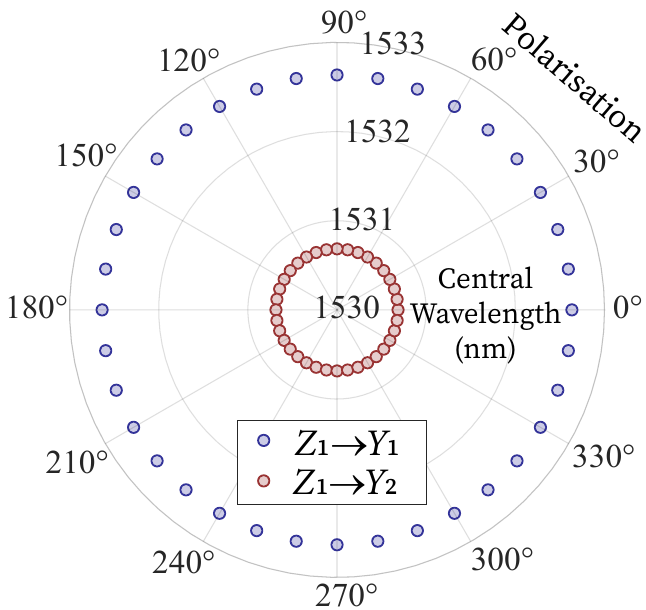}
        \caption{Central wavelength}
    \end{subfigure}%
    ~
    \begin{subfigure}[t]{0.33\textwidth}
        \centering
        \includegraphics[width=\linewidth]{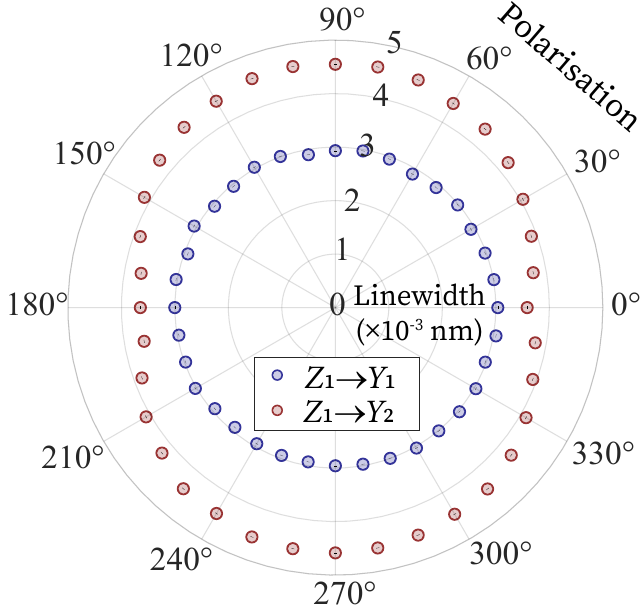}
        \caption{Linewidth}
    \end{subfigure}
    \caption{\label{fig:Polarisation_c}Absorption, central wavelength, and linewidth \JRS{in the $\textbf{k}||a$ configuration} 
    as a function of polarisation angle. $0^\circ$ indicates horizontally polarised light with $\epsilon||c$.}
\end{figure}

The effects of varying the polarisation are shown in Fig.~\ref{fig:Polarisation_a} for the \JRS{$\textbf{k}||c$ configuration and Fig.~\ref{fig:Polarisation_c} for the $\textbf{k}||a$ configuration}. In addition to the absorption, the central wavelengths and linewidths of the $Z_1\to Y_1$ and $Z_1\to Y_2$ transitions are also investigated,

As expected, the central wavelength for both \JRS{configurations} is centred around approximately 1532.6\,nm for the $Z_1\to Y_1$ transition and 1530.8\,nm for the $Z_1\to Y_2$ transition, neither transition exhibiting significant changes with polarisation. Since the electronic levels and the corresponding optical transitions are independent of the polarisation states, the central wavelength at these transitions should be invariant with respect to the incident beam linear polarisation. 

The absorption \JRS{in the $\textbf{k}||c$ configuration} 
for the $Z_1\to Y_1$ transition is approximately 0.700 for all polarisations, and approximately 0.725 for the $Z_1\to Y_2$ transition for all polarisations, with both transitions following a circle with polarisation [Fig.~\ref{fig:Polarisation_a}(a)], implying polarisation independence. However, the absorption polarisation dependence \JRS{in the $\textbf{k}||a$ configuration} 
follows a more interesting result [Fig.~\ref{fig:Polarisation_c}(a)]. The $Z_1\to Y_1$ transition follows an elliptical pattern with polarisation, with maxima when horizontally polarised light is used (where $\epsilon||c$) and minima when vertically polarised light is used (where $\epsilon||a$). Conversely, the absorption \trlee{shows greater} polarisation dependence in the $Z_1\to Y_2$ transition, which follows a \JRS{more} elongated ellipse with polarisation, with maxima when vertically polarised light ($\epsilon||a$) is used and minima when horizontally polarised light is used ($\epsilon||c$). 

The absorption results in this configuration agree qualitatively with previous PL and PLE measurements \cite{becker2024spectroscopic}. Indeed, the $Z_1\to Y_1$ transition involves a change in irreducible representation of the eigenstates $\left[(\Gamma_5+\Gamma_6) \to (\Gamma_7+\Gamma_8)\right]$ whereas the $Z_1\to Y_2$ transition does not $\left[(\Gamma_5+\Gamma_6) \to (\Gamma_5+\Gamma_6)\right]$). For the $\textbf{k}||a$ configuration, different selection rules are fulfilled as the incident beam linear polarisation, $\epsilon$, is varied, which gives rise to angular dependence on polarisation and hence maxima and minima absorptions as shown in Fig. \ref{fig:Polarisation_c}(a). On the other hand, for the $\textbf{k}||c$ configuration, the same selection rules are always satisfied as the incident beam linear polarisation varies, which results in a more circular pattern indicating polarisation independence [Fig. \ref{fig:Polarisation_a}(a)].

The linewidths in the $\textbf{k}||c$ configuration are \JRS{wider} for the $Z_1\to Y_2$ transition than the $Z_1\to Y_1$ transition by approximately 0.7 pm, following a fairly circular pattern with polarisation [Fig.~\ref{fig:Polarisation_a}(c)]. The $Z_1\to Y_1$ transition linewidths are also \JRS{narrower} than the $Z_1\to Y_2$ transition linewidths in the $\textbf{k}||a$ configuration [Fig.~ \ref{fig:Polarisation_c}(c)]. This indicates that the $Z_1\to Y_1$ transition have longer lifetimes, which may translate to longer storage times for quantum states, due to its narrower linewidths in both crystal configurations. However, the $Z_1\to Y_1$ transition linewidths follow a fairly circular pattern with polarisation, while the $Z_1\to Y_2$ transition linewidths have maxima at 90\textdegree~and 270\textdegree~(vertically polarised light, $\epsilon||a$) and minima at 0\textdegree~and 180\textdegree~(horizontally polarised light, $\epsilon||c$). The polarisation dependence in the $Z_1\to Y_2$ transition could be linked to the change in selection rules that are satisfied \cite{becker2024spectroscopic}. 

\subsection{Temperature}
\begin{figure}[htp!]
    \centering
    \begin{subfigure}[t]{0.31\textwidth}
        \centering
        \includegraphics[width=\linewidth]{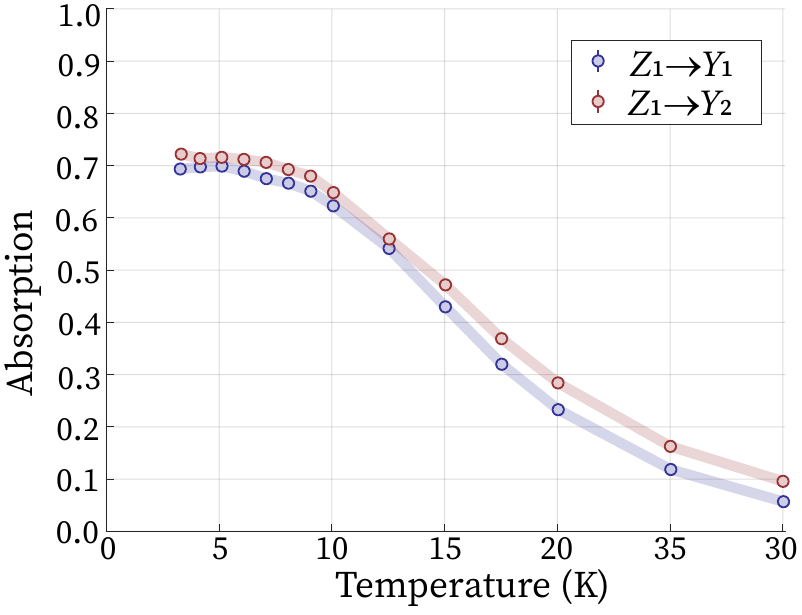}
        \caption{Absorption}
    \end{subfigure}%
    ~ 
    \begin{subfigure}[t]{0.36\textwidth}
        \centering
        \includegraphics[width=\linewidth]{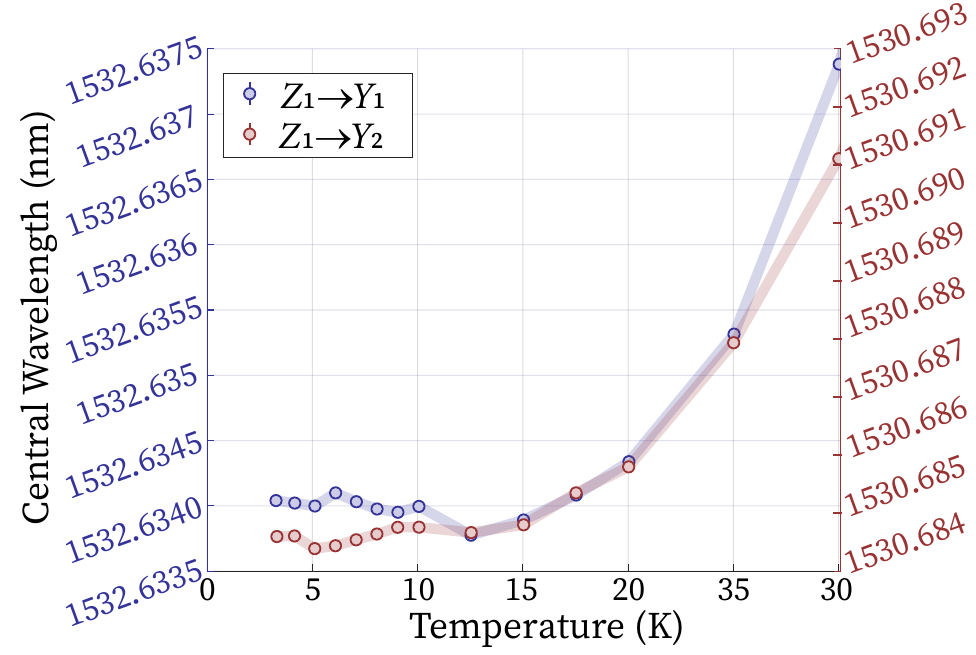}
        \caption{Central wavelength}
    \end{subfigure}%
    ~
    \begin{subfigure}[t]{0.31\textwidth}
        \centering
        \includegraphics[width=\linewidth]{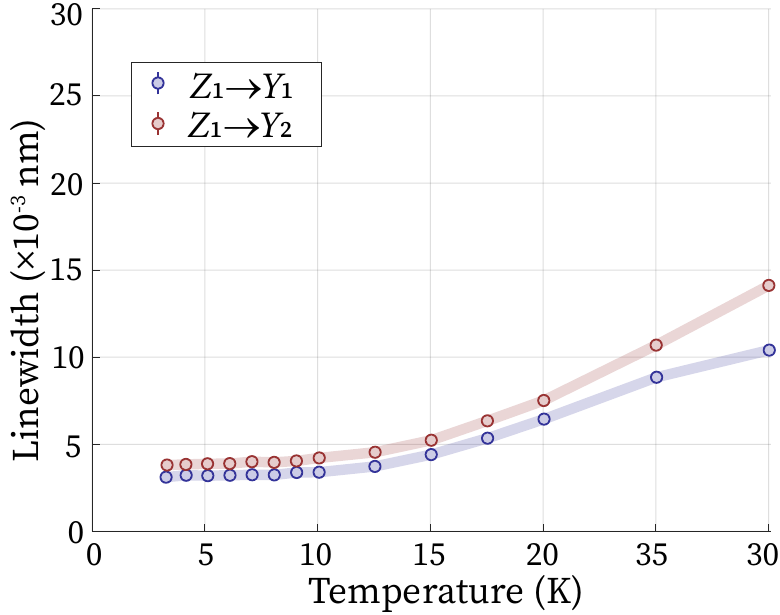}
        \caption{Linewidth}
    \end{subfigure}
    \caption{\label{fig:Temperature_a} Absorption, central wavelength, and linewidth in the 
    $\textbf{k}||c$ configuration with $\epsilon||a$ as a function of temperature.}
\end{figure}

\begin{figure}[htp!]
    \centering
    \begin{subfigure}[t]{0.31\textwidth}
        \centering
        \includegraphics[width=\linewidth]{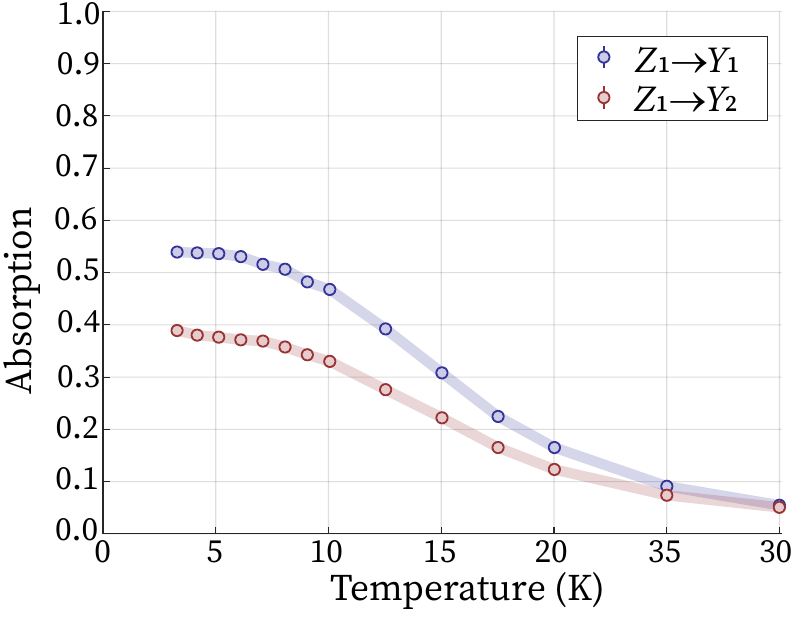}
        \caption{Absorption}
    \end{subfigure}%
    ~ 
    \begin{subfigure}[t]{0.36\textwidth}
        \centering
        \includegraphics[width=\linewidth]{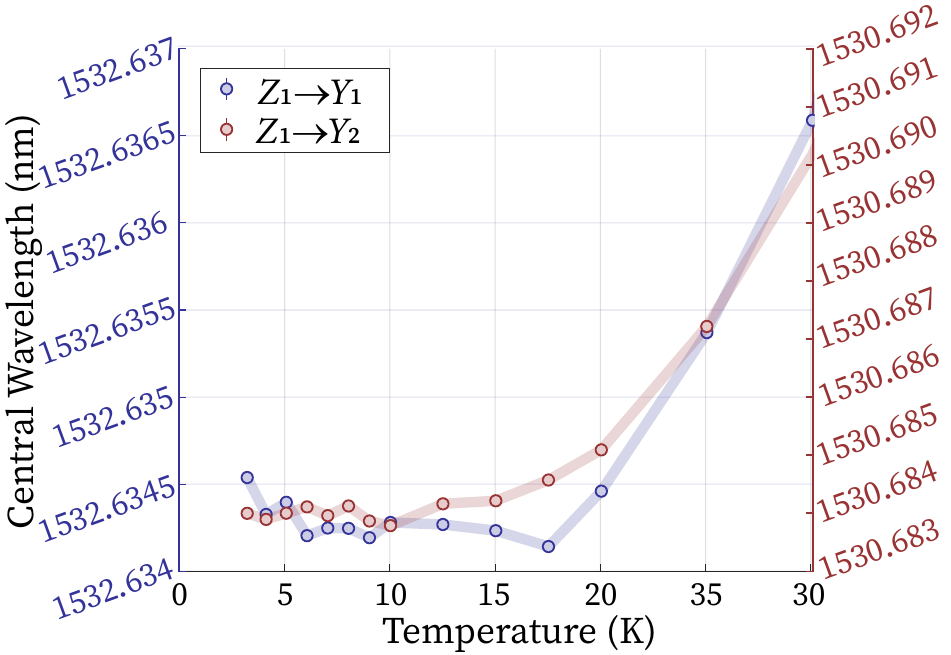}
        \caption{Central wavelength}
    \end{subfigure}%
    ~
    \begin{subfigure}[t]{0.31\textwidth}
        \centering
        \includegraphics[width=\linewidth]{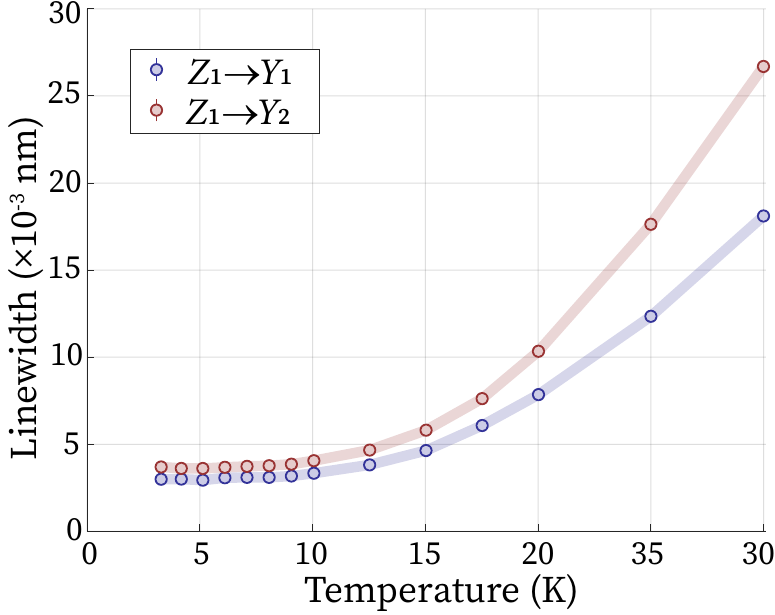}
        \caption{Linewidth}
    \end{subfigure}
    \caption{\label{fig:Temperature_c}Absorption, central wavelength, and linewidth in the 
    $\textbf{k}||a$ configuration with $\epsilon||c$ as a function of temperature.}
\end{figure}

The effects of varying the temperature are shown in Fig.~\ref{fig:Temperature_a} for the 
$\textbf{k}||c$ configuration with $\epsilon||a$ and Fig.~\ref{fig:Temperature_c} for the 
$\textbf{k}||a$ configuration with $\epsilon||c$, respectively. Note that the temperature range is set between 3.2\,K, the base temperature of the Montana cryostat, and 30\,K where significant spectral broadening was observed which causes the absorption dips to disappear. 

It can be seen that increasing the temperature above approximately 15\,K significantly increases the central wavelength for both the $Z_1\to Y_1$ and $Z_1\to Y_2$ transitions in both crystal configurations [Fig. \ref{fig:Temperature_a}(b) and \ref{fig:Temperature_c}(b)]. Below 15\,K, the transitions settle into their respective wavelengths, 1532.634\,nm for the $Z_1\to Y_1$ transition and 1530.684\,nm for the $Z_1\to Y_2$ transition. 

Overall, in both configurations, the $Z_1\to Y_2$ transition linewidth is larger than the $Z_1\to Y_1$ transition linewidth [Fig. \ref{fig:Temperature_a}(c) and \ref{fig:Temperature_c}(c)]. The linewidths for both transitions also increase with temperature.

The absorptions in both transitions begin to decrease approximately above 5\,K. In the 
$\textbf{k}||c$ configuration with $\epsilon||a$, the $Z_1\to Y_2$ transition has a slightly larger absorption than the $Z_1\to Y_1$ transition [Fig.~\ref{fig:Temperature_a}(a)]. Conversely, in the 
$\textbf{k}||a$ configuration with $\epsilon||c$, the $Z_1\to Y_1$ transition has a significantly larger absorption than the $Z_1\to Y_2$ transition [Fig \ref{fig:Temperature_c}(a)]. It is only at  $\sim$30\,K and above that both the transitions have the same near-zero absorption.

As expected, the temperature dependence of Er$^{3+}$:CaWO$_4$ is consistent with previous results in REID materials \cite{myoung2021effects,das2012optical} in which the photoluminescence peaks (absorption dips in this work) decreases in intensity as the temperature increases. Consequently, as these peak (dip) intensity decreases, the spectra broadens causing the measured linewidths to increase. It is also observed, like in \cite{das2012optical}, that the central wavelength of the peaks (dips) slightly increases as the temperature increases from cryogenic temperatures. The spectral broadening is caused by increased thermal motion of the atoms and ions in the crystal which consequently causes decoherence; hence the need for cryogenic temperatures for using REID crystals as quantum memories.

\newcommand*\rot{\rotatebox{90}}
\newcommand*\OK{\ding{51}}

\newpage
\subsection{Qualitative Analysis}
A qualitative comparison has been tabulated in Table \ref{tab:qualitative_analysis} based on the results. It shows the dependency of the absorption, central wavelength, and linewidth based on the polarisation and temperature, as well as the particular crystal axis and transition.

\begin{table}[H]
\centering
\caption{\label{tab:qualitative_analysis}Dependency of absorption, central wavelength, and linewidth on the polarisation and temperature}
\scalebox{0.9}{
\centering
\begin{tabular}{@{}ccclll@{}}
\toprule
Dependency                                                                     & Configuration               & Transition                                                                      & \multicolumn{1}{c}{Polarisation}                                                                                                                & \multicolumn{1}{c}{Temperature}                                                                 \\ \midrule
\multirow{4}{*}{Absorption}                                                    & \multirow{2}{*}{$\textbf{k}||c$} & \multicolumn{1}{c|}{$Z_1$$\to$$Y_1$}                  & \begin{tabular}[c]{@{}l@{}} Sightly smaller \end{tabular}                                                                         & \begin{tabular}[c]{@{}l@{}}Smaller, decreases \\ with temperature\end{tabular}                  \\
                                                                               &                    & \multicolumn{1}{c|}{$Z_1$$\to$$Y_2$}            & \begin{tabular}[c]{@{}l@{}}Slightly larger\end{tabular}                                                                          & \begin{tabular}[c]{@{}l@{}}Larger, decreases \\ with temperature\end{tabular}                   \\
                                                                               & \multirow{2}{*}{$\textbf{k}||a$} & \multicolumn{1}{c|}{$Z_1$$\to$$Y_1$}   & \begin{tabular}[c]{@{}l@{}}Overall smaller, maxima \\ with  horizontally polarised \\ light, minima  with vertically \\ polarised light\end{tabular} & \begin{tabular}[c]{@{}l@{}}Significantly larger, \\ decreases with \\ temperature\end{tabular}  \\
                                                                               &                    & \multicolumn{1}{c|}{$Z_1$$\to$$Y_2$}  & \begin{tabular}[c]{@{}l@{}}Overall larger, maxima \\ with  vertically polarised \\ light, minima  with \\ horizontally polarised light\end{tabular}  & \begin{tabular}[c]{@{}l@{}}Significantly smaller, \\ decreases with \\ temperature\end{tabular} \\
\multirow{4}{*}{\begin{tabular}[c]{@{}c@{}}Central \\ Wavelength\end{tabular}} & \multirow{2}{*}{$\textbf{k}||c$} & \multicolumn{1}{c|}{$Z_1$$\to$$Y_1$}                        & \begin{tabular}[c]{@{}l@{}}No significant \\ changes\end{tabular}                                                                               & \begin{tabular}[c]{@{}l@{}}Larger, increases \\ with temperature\end{tabular}                   \\
                                                                               &                    & \multicolumn{1}{c|}{$Z_1$$\to$$Y_2$}        & \begin{tabular}[c]{@{}l@{}}No significant \\ changes\end{tabular}              & \begin{tabular}[c]{@{}l@{}}Smaller, increases \\ with temperature\end{tabular}                  \\
                                                                               & \multirow{2}{*}{$\textbf{k}||a$} & \multicolumn{1}{c|}{$Z_1$$\to$$Y_1$}                       & \begin{tabular}[c]{@{}l@{}}No significant \\ changes\end{tabular}                                                                               & \begin{tabular}[c]{@{}l@{}}Smaller, increases \\ with temperature\end{tabular}                  \\
                                                                               &                    & \multicolumn{1}{c|}{$Z_1$$\to$$Y_2$}         & \begin{tabular}[c]{@{}l@{}}No significant \\ changes\end{tabular}          & \begin{tabular}[c]{@{}l@{}}Larger, increases \\ with temperature\end{tabular}                   \\
\multirow{4}{*}{Linewidth}                                                     & \multirow{2}{*}{$\textbf{k}||c$} & \multicolumn{1}{c|}{$Z_1$$\to$$Y_1$}                 & \begin{tabular}[c]{@{}l@{}}Significantly smaller\end{tabular}                                                           & \begin{tabular}[c]{@{}l@{}}Smaller, increases \\ with temperature\end{tabular}                  \\
                                                                               &                    & \multicolumn{1}{c|}{$Z_1$$\to$$Y_2$}    & \begin{tabular}[c]{@{}l@{}}Significantly larger\end{tabular}           & \begin{tabular}[c]{@{}l@{}}Larger, increases \\ with temperature\end{tabular}                   \\
                                                                               & \multirow{2}{*}{$\textbf{k}||a$} & \multicolumn{1}{c|}{$Z_1$$\to$$Y_1$}         & \begin{tabular}[c]{@{}l@{}}Smaller, constant with\\ polarisation\end{tabular}     & \begin{tabular}[c]{@{}l@{}}Smaller, increases \\ with temperature\end{tabular}                  \\
                                                                               &                    & \multicolumn{1}{c|}{$Z_1$$\to$$Y_2$}         & \begin{tabular}[c]{@{}l@{}}Larger, maxima with\\ vertically polarised  light,
                                                    \\minima with horizontally\\ polarised light\end{tabular}             & \begin{tabular}[c]{@{}l@{}}Larger, increases \\ with temperature\end{tabular}                   \\ \bottomrule
\end{tabular}
}
\end{table}

\section{Discussion} 
The experimental characterization of the $Z_1 \rightarrow Y_1$ and $Z_1\rightarrow Y_2$ optical transitions in $\mathrm{Er^{3+}:CaWO_4}$ provides insight into their suitability for quantum memory applications. Among the two transitions, the $Z_1 \rightarrow Y_1$ transition exhibits narrower linewidth, higher absorption, and polarization insensitivity compared to the $Z_1\rightarrow Y_2$ transition. These features make the $Z_1 \rightarrow Y_1$ transition more suitable for quantum memory protocols such as the atomic frequency comb (AFC) \cite{afzelius2009multimode,bonarota2010efficiency,afzelius2010demonstration}, where higher absorption is favorable for higher efficiency. Furthermore, the polarization-insensitive absorption characteristics makes the $Z_1 \rightarrow Y_1$ transition suitable for quantum memories designed for storing photonic polarization qubits where arbitrary polarization states can be stored with high fidelity.

The linewidth of the $Z_1\rightarrow Y_2$ transition is larger than the linewidth of the $Z_1 \rightarrow Y_1$. The measurements also showed that both linewidth decreases with lower temperature and stops decreasing below approximately 5~K. This behavior is commonly attributed to the suppression of phonon-induced decoherence \cite{budoyo2018phonon}. In general, the narrower linewidth, therefore, tends to translates to longer storage times. However, it is important to recognize that the linewidth measurements reflect inhomogeneous broadening, and does not directly inform about the homogenous linewidth, which ultimately determines the optical coherence time and hence the achievable storage time.

Although the temperature dependence measurements showed the saturation behavior  below approximately 5~K, it would be interesting to explore the phonon-induced decoherence effects near absolute zero -- at temperatures as low as 20~mK, it is known that phonons arising from lattice vibrations in the crystal causes decoherence \cite{budoyo2018phonon}. The homogeneous linewidth is strongly temperature-dependent, with different phonon-induced decoherence mechanisms dominating at different temperature ranges. The precise characterization through coherent spectroscopy and mitigation of this phenomenon could be an avenue for future work.

Although the absorption, central wavelength, and linewidth differ depending on the \trlee{electronic} transition, crystal axis \trlee{orientation relative to the polarisation of incident light}, the use of the $Z_1\to Y_1$ transition with the laser oriented along the crystal $c$-axis 
is recommended. It has a smaller linewidth than the $Z_1\to Y_2$ transition, along with a constant central wavelength (for temperatures below 5\,K). Additionally, it has an overall greater absorption than the configuration with the laser oriented along the crystal $a$-axis and with $\epsilon||c$. 
Significantly, it is much less polarisation dependent compared to the $\textbf{k}||a$ configuration. 
In terms of engineering an Er$^{3+}$:CaWO$_4$-based quantum memory, a cryostat must be used to keep the temperature below 5\,K. In addition, different sources of noise must be characterised and the quantum memory designed to mitigate them. 




\section{Conclusion} 
The polarisation and temperature dependence of the absorption spectra of Er$^{3+}$:CaWO$_4$ was studied in the endeavour of developing the solid-state rare-earth ion-doped crystal as a quantum memory. The results show that the central wavelength of the $Z_1\to Y_1$ and $Z_1\to Y_2$ transitions is independent of polarisation and temperature at cryogenic temperatures (below 5~K). In addition, the linewidth of the $Z_1\to Y_1$ transition is overall smaller making it more favourable for long storage times than the $Z_1\to Y_2$ transition. Overall, the 
$Z_1\to Y_1$ transition saturates slower than the $Z_1\to Y_2$ making it more favourable for longer storage times. Significantly, the $\textbf{k}||c$ configuration absorption is polarisation independent, making it the more favourable Er$^{3+}$:CaWO$_4$ configuration to store quantum states as the storage times will be fairly constant for any polarisation.


\paragraph{\textbf{Acknowledgments}}
This research is supported by A*STAR under Project No. C230917009, Q.InC Strategic Research and Translational Thrust; the MTC Young Investigator Research Grant (Award \# M24N8c0110); CQT++ Core Research Funding Grant (A*STAR); 
\NI{
and Grant-in-Aid for Scientific Research (Grant No. 22K03507) from the Japan Society for the Promotion of Science.
}

\paragraph{\textbf{Conflict of Interest}}
The authors declare no conflicts of interest.

\section*{References}
\bibliographystyle{IEEEtran}
\bibliography{References}

\end{document}


\begin{center}
    \large \textbf{Supplementary Material}\\[0.5em]
    \small for the article: \textit{“Polarisation and Temperature Dependence of Er$^{3+}$:CaWO$_4$ -- Towards a solid-state rare-earth ion doped quantum memory”}
\end{center}



\section{Crystal Field Model}
\label{Sec:CF}
The theory of the crystal field for the $4f$ electron systems in Er$^{3+}$ is described here.
Suppose that the crystal field can be described with a potential acting on the $4f$ orbitals \cite{Abragam1970}: 
\begin{align}
 \hat{h}_\text{CF} &= \sum_{kq} b_{kq} Y_{kq}(\theta, \phi),
 \label{Eq:hCF}
\end{align}
where $b_{kq}$ are the single electron crystal field parameters, and $Y_{kq}$ are the spherical harmonics acting on the $4f$ orbitals. 
The $b_{kq}$ comes from various effects such as chemical bond between the $4f$ and ligand atoms \cite{Ungur2017}, 
traditional Coulomb repulsion between the $4f$ electrons and ligand electrons, etc.

For the $N$ electronic states, the crystal-field Hamiltonian is 
\begin{align}
 \hat{H}_\text{CF} &= \sum_{mm'\sigma} \sum_{kq} b_{kq} \langle fm| Y_{kq} |fm'\rangle \hat{f}_{m\sigma}^\dagger \hat{f}_{m'\sigma}
 \label{Eq:HCF0}
\end{align}
with the $4f$ electron creation and annihilation operators, $\hat{f}_{m\sigma}^\dagger$ and $\hat{f}_{m\sigma}$. 
Using the Wigner-Eckart theorem,
\begin{align}
 \hat{H}_\text{CF}
 &= \sum_{mm'\sigma} \sum_{kq} b_{kq} \frac{\langle f\Vert Y_{k} \Vert f \rangle}{\sqrt{[l_f]}} (l_f m|l_f m', kq) \hat{f}_{m\sigma}^\dagger \hat{f}_{m'\sigma}
 \\
 &= \sum_{mm'\sigma} \sum_{kq} b_{kq} \frac{\langle f\Vert Y_{k} \Vert f \rangle}{\sqrt{[l_f]}} 
 (-1)^{l_f - m'} \sqrt{\frac{[l_f]}{[k]}} 
 \nonumber\\
 &\times
 (kq|l_f m, l_f -m')
 \hat{f}_{m\sigma}^\dagger \hat{f}_{m'\sigma}.
 \label{Eq:HCF1}
\end{align}
Here $(c\gamma|a\alpha, b\beta)$ are the Clebsch-Gordan coefficients of Condon-Shortley's convention,
$[k] = 2k+1$, and $\langle f\Vert Y_{k} \Vert f \rangle$ the reduced matrix elements. 
By using Racah's unit operator \cite{Judd1967}, 
\begin{align}
 \sum_{\sigma} \hat{f}_{m\sigma}^\dagger \hat{f}_{m'\sigma}
 &=
 (-1)^{l_f-m'} \sum_{kq} (kq|l_f m, l_f -m') \hat{U}^{(k)}_q,
 \label{Eq:U}
\end{align}
the crystal field Hamiltonian (\ref{Eq:HCF0}) reduces to 
\begin{align}
 \hat{H}_\text{CF} &= \sum_{kq} b_{kq}  \frac{\langle f\Vert Y_{k} \Vert f \rangle}{\sqrt{[k]}} \hat{U}^{(k)}_q.
 \label{Eq:HCFL}
\end{align}
So far, the crystal-field Hamiltonian acts on the entire electronic states originating from the $4f^N$ configurations. 

For the analysis of the low-energy electronic states, the crystal-field Hamiltonian is projected into ${}^4I_J$ multiplet states. 
The low-lying $J$ multiplet states, in a good approximation, originate from the lowest $LS$ term states:
\begin{align}
 |JM_J\rangle &= \sum_{M_LM_S} |LM_L, SM_S\rangle (LM_L, SM_S|JM_J)
\end{align}
To obtain the crystal field Hamiltonian within single $J$ multiplet states, the unit operator must first be projected into a single $LS$ term:
\begin{align}
 \hat{U}^{(k)}_q &\rightarrow \frac{\langle {}^{[S]}L\Vert \hat{U}^{(k)} \Vert {}^{[S]}L \rangle}{\sqrt{[k]}} 
 \sum_{M_L M_L'} (-1)^{L-M_L'} 
 \nonumber\\
 &\times
 (kq|LM_L, L-M_L') |LM_L\rangle \langle LM_L'| \otimes \hat{1}_S.
\end{align}
Then, the unit operator is projected into single $J$ multiplet states: 
\begin{align}
 \hat{U}^{(k)}_q &\rightarrow 
 \sum_{M_J M_J'}
 \frac{\langle {}^{[S]}L\Vert \hat{U}^{(k)} \Vert {}^{[S]}L \rangle}{\sqrt{[k]}} 
 \sum_{M_L M_L'} (-1)^{L-M_L'} 
 \sum_{M_S}
 \nonumber\\
 &\times
 (kq|LM_L, L-M_L') 
 (JM_J|LM_L, SM_S)
 \nonumber\\
 &\times
 (JM_J'|LM_L, SM_S)
 |JM_J\rangle \langle JM_J'|
 \\
 &=
 \frac{\langle {}^{[S]}L\Vert \hat{U}^{(k)} \Vert {}^{[S]}L \rangle}{\sqrt{[k]}} 
 (-1)^{L+S+J+k} [J] 
 \begin{Bmatrix}
  L & k & L \\
  J & S & J 
 \end{Bmatrix}
 \nonumber\\
 &\times
 \sum_{M_J M_J'} (-1)^{J-M_J'} (kq|JM_J, J-M_J') |JM_J\rangle \langle JM_J'|.
\end{align}
By defining the irreducible tensor operators (ITOs) acting on the $J$ multiplet states \cite{Santini2009} as 
\begin{align}
 \hat{\mathcal{T}}_{kq}^J &= \sum_{MM'} (-1)^{J-M'} (kq|JM, J-M') |JM\rangle \langle JM'|,
 \label{Eq:Tkq}
\end{align}
the Racah's unit operator within the same space is 
\begin{align}
 \hat{U}^{(k)}_q &= 
 \frac{\langle {}^{[S]}L\Vert \hat{U}^{(k)} \Vert {}^{[S]}L \rangle}{\sqrt{[k]}} 
 (-1)^{L+S+J+k} [J] 
 \begin{Bmatrix}
  L & k & L \\
  J & S & J 
 \end{Bmatrix}
 \hat{\mathcal{T}}_{kq}^J.
\end{align}
With this expression, the crystal field Hamiltonian within single $J$ multiplet states is 
\begin{align}
 \hat{H}_\text{CF} &= \sum_{kq} B_{kq}^J \hat{\mathcal{T}}_{kq}^J, 
 \label{Eq:HCFJ}
 \\
 B_{kq}^J &= b_{kq}  \frac{\langle f\Vert Y_{k} \Vert f \rangle}{\sqrt{[k]}}  \frac{\langle {}^{[S]}L\Vert \hat{U}^{(k)} \Vert {}^{[S]}L \rangle}{\sqrt{[k]}} 
 \nonumber\\
 &\times
 (-1)^{L+S+J+k} [J] 
 \begin{Bmatrix}
  L & k & L \\
  J & S & J 
 \end{Bmatrix}.
 \label{Eq:BCFJ}
\end{align}
The crystal field parameters (\ref{Eq:BCFJ}) for different $J$'s have the following relation:
\begin{align}
 \frac{B_{kq}^J}{B_{kq}^{J'}} &= (-1)^{J-J'} \frac{[J]}{[J']}
 \frac{ \begin{Bmatrix}
  L & k & L \\
  J & S & J 
 \end{Bmatrix}}{
 \begin{Bmatrix}
  L & k & L \\
  J' & S & J' 
 \end{Bmatrix}}.
 \label{Eq:BCFJ_rel}
\end{align}

Now, the above framework shall be applied to the Er impurity in CaWO$_4$. 
It is assumed that the Er site has $S_4$ point group symmetry, although, according to the experimental structure (your cif file), the Er site has slightly lower symmetry than the $S_4$ symmetry. 
The crystal-field mixing between different $J$ multiplet states was ignored. 
Within the above approximations, the crystal field Hamiltonian within the ${}^4I_{J}$ multiplet states is 
\begin{align}
 \hat{H}^J_\text{CF} &= 
 B_{20}^J \hat{\mathcal{T}}_{20}^J + 
 B_{40}^J \hat{\mathcal{T}}_{40}^J + 
 B_{44}^J \left( \hat{\mathcal{T}}_{44}^J + \hat{\mathcal{T}}_{4,-4}^J \right)
 \nonumber\\
 &+
 B_{60}^J \hat{\mathcal{T}}_{60}^J + 
 B_{64}^J \left( \hat{\mathcal{T}}_{64}^J + \hat{\mathcal{T}}_{6,-4}^J \right).
 \label{Eq:HCF_152}
\end{align}
Up to 6th-rank terms are considered, ignoring the higher terms arising from, for example, $J$ mixing. 
Because of the symmetry, the nonzero crystal field coefficients are $B_{20}$, $B_{40}$, $B_{44}$, $B_{60}$, and $B_{64}$.

Within the present formalism, the crystal field parameters for $J=15/2$ and $J=13/2$ of Er$^{3+}$ ion satisfy Eq. (\ref{Eq:BCFJ_rel}):
\begin{align}
 \frac{B_{20}^{J=13/2}}{B_{20}^{J=15/2}} &= \frac{21}{5\sqrt{19}} = 0.964,
 \label{Eq:ratio_2}
 \\
 \frac{B_{4q}^{J=13/2}}{B_{4q}^{J=15/2}} &= 7 \sqrt{\frac{2}{209}} = 0.685,
 \label{Eq:ratio_4}
 \\
 \frac{B_{6q}^{J=13/2}}{B_{6q}^{J=15/2}} &= \frac{7}{2} \sqrt{\frac{7}{1045}} = 0.286.
 \label{Eq:ratio_6}
\end{align}
In real materials, the ratios could vary due to the ignored effects, such as the energy dependence of the covalency \cite{Ungur2022, Dergachev2025}, while the variations would not be significant. 
The relations will be used to check the validity of the fitting. 

\newpage

\section{Results from the Crystal Field Model}
The experimental energy levels of Er$^{3+}$ impurity in CaWO$_4$ \cite{Enrique1971, Becker2024} are fitted to the crystal field model. 
Table \ref{Table:CF} shows the crystal-field parameters and crystal-field levels.
The $J=15/2$ ($Z_1$-$Z_8$) and $J=13/2$ ($Y_1$-$Y_7$) crystal field levels are fit separately. 
The ratios of $B^{13/2}/B^{15/2}$ are as follows. 
\begin{align*}
 \frac{B_{20}^{13/2}}{B_{20}^{15/2}} &= 1.075,
 \\
 \frac{B_{40}^{13/2}}{B_{40}^{15/2}} &= 0.83, \quad 
 \frac{B_{44}^{13/2}}{B_{44}^{15/2}} = 0.66,
 \\
 \frac{B_{60}^{13/2}}{B_{60}^{15/2}} &= -0.23, \quad
 \frac{B_{60}^{13/2}}{B_{60}^{15/2}} = 0.34.
\end{align*}
Except for the $(6,0)$ component with small crystal field parameters, the ratios are in line with those within the simple model, Eqs. (\ref{Eq:ratio_2})-(\ref{Eq:ratio_6}). The 

\begin{table}[htbp]
\captionsetup{justification=raggedright,singlelinecheck=false}
\caption{Crystal-field parameters (cm$^{-1}$). The lowest $Z$ and $Y$ levels are set to zero for the comparison with the experimental data.}
\begin{center}
\label{Table:CF}
\begin{tabular}{lllllll}
         & \multicolumn{2}{c}{Ref. \cite{Enrique1971}} & \multicolumn{4}{c}{Ref. \cite{Becker2024} (data 1)} \\
         & \multicolumn{2}{c}{$Z$}                     & \multicolumn{2}{c}{$Z$} & \multicolumn{2}{c}{$Y$} \\
         & calc. & exp.   & calc. & exp.   & calc. & exp.  \\
         \hline
$B_{20}$ &   133.0  & &   119.6  & &   128.6  \\ 
$B_{40}$ & $-164.0$ & & $-146.1$ & & $-120.7$ \\
$B_{44}$ &   186.0  & &   187.6  & &   123.2  \\
$B_{60}$ &   $-4.8$ & &   $-6.5$ & &     1.5  \\
$B_{64}$ &   281.8  & &   284.3  & &    96.2  \\
\hline
\multicolumn{7}{c}{Crystal field levels (meV)} \\
       & calc.  & exp.  & calc. & exp.   & calc. & exp.  \\
$1$    &  0     &   0   &   0    &  0    &   0    &     0  \\
$2$    &  19.63 &  19.2 & 19.77 & 20.13 &  10.05 &   8.32 \\ 
$3$    &  21.52 &  24.9 & 22.82 & 25.84 &  49.23 &  48.59 \\ 
$4$    &  46.07 &  51.1 & 48.49 & 51.62 & 129.64 & 130.25 \\ 
$5$    & 228.36 & 227.9 & 225.8 & 227.7 & 159.69 & 158.67 \\ 
$6$    & 269.11 & 265.9 & 269.3 & 269.5 & 179.17 & 176.45 \\ 
$7$    & 293.78 & -     & 291.2 & 293.8 & 193.11 & 193.22 \\ 
$8$    & 318.79 & 319.4 & 317.4 & 318.3 &        &        \\
\end{tabular}
\end{center}
\end{table}
\FloatBarrier  


\newpage

\section{Experimental Methods: Data Extraction and Analysis}
The oscilloscope used for data collection was set to Normal acquisition, with a sampling rate of 2~MSa/s. To acquire the data, the laser's wavelength was varied at a speed $v_{\mathrm{sweep}}$ of 1~nm/s, its lowest value, for maximum wavelength precision. At the beginning of a sweep, the laser sends an electrical trigger to the oscilloscope, which is set to Single trigger mode. This begins the plotting of the photodetector's output voltage, which is directly proportional to the optical input power detected. By setting the time of the electrical trigger as $t=0$, the sweep time was converted to wavelength by utilising the linear relationship between them, 

\begin{equation}
    \lambda(t) = \lambda_{\mathrm{start}} + v_{\mathrm{sweep}}t
\end{equation}
where $\lambda_{\mathrm{start}}$ is the start wavelength for a particular wavelength range.

For every characterisation sweep performed, nine more repeat sweeps were also performed. These ten sweeps were then linearly interpolated by oversampling the raw data with $10\times10^{6}$ points to an averaged plot with a single ``common" wavelength axis. 

The instruments used were connected to a central computer via a router. Using Python SCPI commands, the laser powers, wavelengths, and the cryostat temperatures could be controlled on the computer. To ensure that the data captured by the oscilloscope was not getting clipped, while also achieving high resolution, some preliminary oscilloscope settings were determined and hard-coded into the oscilloscope under the necessary experimental conditions.

The interpolated absorption dip signal was transformed by

\begin{equation}
    y = -(1 - \mu),
\end{equation}
where $\mu$ is the mean voltage (excluding the absorption dip), and a Gaussian fit was used using MATLAB R2022a, which takes the form

\begin{equation}
    f(x) = ae^{-(\frac{x-b}{c})^2}.
\end{equation}

The corresponding errors in $a,b$, and $c$ were calculated based on a 95\% confidence interval. These errors have been propagated in calculations using $a,b$, and $c$. From this Gaussian fit, the absorption, central wavelength, and linewidth could be extracted. The absorption was calculated as

\begin{equation}
    A = 1 - \frac{\mu - a}{\mu},
\end{equation}
representing the percentage of absorption.

The central wavelength was taken as $b$ from the fit. The linewidth was calculated as 

\begin{equation}
    \mathrm{LW} = \sqrt{2}c.
\end{equation}

\appendix
\section{Comments on ITOs}
ITOs (see e.g. Ref. \cite{Santini2009}) were used rather than traditional Steven's operators. 
\begin{itemize}
 \item ITOs are proportional to Steven's operators. 
 \item Constructing the ITOs is much easier than Steven's operators. For the ITOs, only Clebsch-Gordan coefficients need to be corrected, which is easily accomplished. On the other hand, the lists of Steven's operators often contain errors. 
 \item With ITOs, symmetry properties (Wigner-Eckart theorem) can be used, which is not the case with Steven's operators.
 \item The ITOs fulfil the orthonormality:
 \begin{align}
  \text{Tr}\left[ \hat{\mathcal{T}}_{kq}^\dagger \hat{\mathcal{T}}_{k'q'} \right] = \delta_{kk'} \delta_{qq'}.
 \end{align}
 The trace is over the $J$ multiplet defining the ITOs.
\end{itemize}

\section*{References}
\bibliographystyle{IEEEtran}
\bibliography{ref}